\documentclass[aps,reprint,article,prx]{revtex4-1}
\usepackage{amsmath}
\usepackage{dcolumn}
\usepackage{graphicx}
\usepackage{float}
\usepackage{color}
\usepackage{booktabs}
\usepackage{multirow}

\begin{document}

\title{Superconductivity and vortex structure on Bi$_{2}$Te$_{3}$/FeTe$_{0.55}$Se$_{0.45}$ heterostructures with different thickness of Bi$_{2}$Te$_{3}$ films}

\author{Kailun Chen, Mingyang Chen, Chuanhao Wen, Zhiyong Hou, Huan Yang,$^*$ and Hai-Hu Wen$^\dag$}

\affiliation{National Laboratory of Solid State Microstructures and Department of Physics, Collaborative Innovation Center of Advanced Microstructures, Nanjing University, Nanjing 210093, China}

\begin{abstract}
Using scanning tunnel microscopy (STM), we investigate the superconductivity and vortex properties in topological insulator Bi$_{2}$Te$_{3}$ thin films grown on the iron-based superconductor FeTe$_{0.55}$Se$_{0.45}$. The proximity-induced superconductivity weakens in the Bi$_{2}$Te$_{3}$ film when the thickness of the film increases. Unlike the elongated shape of vortex cores observed in the Bi$_{2}$Te$_{3}$ film with 2-quintuple-layer (QL) thickness, the isolated vortex cores exhibit a star shape with six rays in the 1-QL film, and the rays are along the crystalline axes of the film. This is consistent with the sixfold rotational symmetry of the film lattice, and the proximity-induced superconductivity is still topologically trivial in the 1-QL film. At a high magnetic field, when the direction between the two nearest neighbored vortices deviates from that of any crystalline axes, two cores connect each other by a pair of adjacent rays, forming a new type of electronic structure of vortex cores. On the 3-QL film, the vortex cores elongate along one of the crystalline axes of the Bi$_{2}$Te$_{3}$ film, similar to the results obtained on 2-QL films. The elongated vortex cores indicate a twofold symmetry of the superconducting gap induced by topological superconductivity with odd parity. This observation confirms possible topological superconductivity in heterostructures with a thickness of more than 2 QLs. Our results provide rich information for the vortex cores and vortex-bound states on the heterostructures consisting of the topological insulator and the iron-based superconductor.
\end{abstract}

\maketitle
\section{Introduction}

Topological superconductors have superconducting (SC) bulk states protected by the SC gap and gapless surface states, and they can host Majorana-bound states which attract considerable attention because of the potential application of topological quantum computation \cite{SCZhangReview,AndoReview}. However, the natural topological superconductors are rare. An effective way to realize topological superconductors is to dope topological insulators. For instance, SC materials $M_{x}$Bi$_{2}$Se$_{3}$ ($M$ = Nb, Sr, or Cu) are proven to have the feature of topological superconductivity \cite{CuBiSe1,CuBiSe2,CuBiSe3,SrBiSe,GDu}. Later, some iron-based superconductors \cite{HuReview,FeTeSeMajorana,FeTeSearpes,LiFeOHFeSe,CaKFe4As4,FeTeSeHanaguri,WangZY,LiFeAs} and some other superconductors \cite{WS2LiW,CVSWangZY,Jiaolin} are regarded as topological superconductors because they have the non-trivial topological band or chiral-triplet topological superconductivity. Another approach to realizing topological superconductors is to induce superconductivity in the topological-insulating (TI) layer from the neighboring superconductor through the proximity effect \cite{LFuProximity}. This heterostructure consisting of the topological insulator and the conventional superconductor can host Majorana zero modes in vortex cores \cite{JiaReview,Jiaprl}. Meanwhile, Majorana modes are observed at the ends of one-dimensional magnetic atom chains \cite{FeChain} or the edge of two-dimensional ferromagnetic islands \cite{Feisland,CrBr3} grown on the $s$-wave superconductors.

To achieve a high critical temperature ($T_\mathrm{c}$) topological superconductivity, high-$T_\mathrm{c}$ superconductors are used to fabricate heterostructures. The proximity-induced superconductivity is detected in Bi$_2$Te$_3$/Bi$_2$Sr$_2$CaCu$_2$O$_{8+\delta}$ heterostuctures \cite{TIonBi2212,TIonBi2212ARPES}. In addition, a twofold symmetric SC gap is observed in such heterostructures \cite{TIonBi2212Wan}, which is consistent with the $\Delta_{4y}$ notation proposed for the topological superconductivity \cite{LFuCuBiSe}. High-$T_\mathrm{c}$ iron-based superconductors can also induce superconductivity in the neighboring TI layer. From our previous research, a clear proximity-induced SC gap with twofold symmetry is obtained in Bi$_2$Te$_3$/FeTe$_{0.55}$Se$_{0.45}$ heterostructures with the Bi$_2$Te$_3$ film thickness of 2 quintuple layers (QLs) \cite{ChenMYSA}, and the elongated vortex cores are also observed on heterostructure. Both of these observations are the evidence of the possible topological superconductivity. When the Bi$_{2}$Te$_{3}$ film is thicker than 4 QLs, the proximity-induced superconductivity is still observed in the TI layer of the Bi$_2$Te$_3$/FeTe$_{0.55}$Se$_{0.45}$ heterostructure \cite{BiTeonFeTeSe}. Besides the heterostructures containing the TI layer, the materials with the strong spin-orbital coupling \cite{Biisland} or the magnetic order \cite{MnTe} grown on the iron-based superconductors can also fabricate the platform to hold Majorana zero mode.

In this work, we report detailed results of SC and vortex properties in Bi$_{2}$Te$_{3}$ thin films with different thicknesses grown on the iron-based superconductor FeTe$_{0.55}$Se$_{0.45}$. Proximity-induced superconductivity is observed in Bi$_2$Te$_3$ films, and it becomes weak with the increase of the film thickness. On the 1-QL Bi$_{2}$Te$_{3}$/FeTe$_{0.55}$Se$_{0.45}$, the isolated vortex cores are sixfold star-shaped, and their rays extend away along the crystalline axes of the film. When the vortex lattice arrangement is distorted at a high magnetic field, a pair of rays becomes bent and connected to the nearest vortices. While in the heterostructures with TI layers thicker than 2 QLs, twofold symmetric vortex cores are observed. We ascribe the proximity-induced superconductivity on films thicker than 2 QLs to the topologically nontrivial one due to the existence of topological surface states.

\section{Experimental methods}

The single crystals of FeTe$_{0.55}$Se$_{0.45}$ were synthesized by the self-flux method \cite{samplegrowth}. The interstitial Fe atoms have been eliminated by annealing the sample at $400^\circ$C for 20 hours in an O$_{2}$ atmosphere, followed by quenching in liquid nitrogen. The single crystal was cleaved in an ultrahigh vacuum with a pressure of about $1\times10^{-10}$ Torr, and then the FeTe$_{0.55}$Se$_{0.45}$ substrates were heated to $280^\circ$C to degas. Bi$_{2}$Te$_{3}$ thin films were grown on the cleaved $ab$-plane surfaces of FeTe$_{0.55}$Se$_{0.45}$ by the molecular beam epitaxy (MBE) technique. High-purity Bi (99.999\%) and Te (99.999\%) powders were evaporated from effusion cells (CreaTec) with a flux ratio of about 1:18 of Bi/Te, and the temperature of the substrate was held at $265^\circ$C during the deposition. The film growth rate was about 0.5 QL/min, and the growth process was monitored by reflection high-energy electron diffraction in the MBE chamber. The thickness of 1 QL is about 1 nm for the Bi$_{2}$Te$_{3}$ film \cite{ChenMYSA}. After the film growth, samples were annealed at 265$^\circ$C for 5 min. The scanning tunneling microscopy/spectroscopy (STM/STS) measurements were carried out in a USM-1300 system (Unisoku Co. Ltd.) with an ultrahigh vacuum, low temperature, and high magnetic field. The tunneling spectra were measured by a lock-in technique with an amplitude of 0.3 mV and a frequency of 938 Hz. All measurements are taken at 0.4 K. The magnetic field was applied along the $c$-axis of the FeTe$_{0.55}$Se$_{0.45}$ substrate or equivalently perpendicular to the Bi$_{2}$Te$_{3}$ film.

\section{Results}

\subsection{Spectroscopic characterization}

\begin{figure}
\includegraphics[width=8cm]{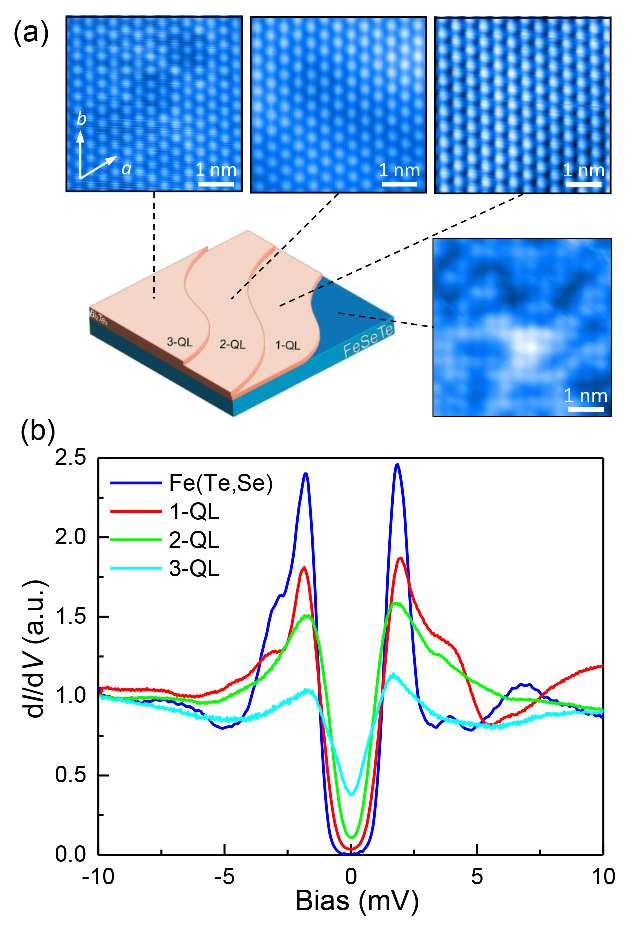}
\caption{(a) Schematic image of Bi$_{2}$Te$_{3}$/FeTe$_{0.55}$Se$_{0.45}$ heterostructures with different thicknesses of Bi$_{2}$Te$_{3}$. The atomically resolved topography is measured on FeTe$_{0.55}$Se$_{0.45}$ substrate and Bi$_2$Te$_3$ films (set-point tunneling conditions: $V_\mathrm{set} = 10$ mV, $I_\mathrm{set} = 100$ pA). (b) Typical tunneling spectra measured in FeTe$_{0.55}$Se$_{0.45}$ and Bi$_2$Te$_3$ films ($V_\mathrm{set} = 20$ mV, $I_\mathrm{set} = 200$ pA).}\label{fig1}
\end{figure}

The structure of Bi$_{2}$Te$_{3}$/FeTe$_{0.55}$Se$_{0.45}$ heterostructure is shown schematically in Fig.~\ref{fig1}(a).  Typical atomically resolved topographic images measured on FeTe$_{0.55}$Se$_{0.45}$ and Bi$_{2}$Te$_{3}$ films are also shown in the figure. The topography of FeTe$_{0.55}$Se$_{0.45}$ is imaged before the film growth, and the surface is the square lattice consisting of Te or Se atoms \cite{ChenMYCdGM} with a lattice constant of about 3.80 \AA. The top surface of the Bi$_{2}$Te$_{3}$ film is the hexagonal lattice of Te atoms with a lattice constant of about 4.38 \AA. Typical tunneling spectra are shown in Fig.~\ref{fig1}(b), and they are measured on FeTe$_{0.55}$Se$_{0.45}$ before the film growth and Bi$_{2}$Te$_{3}$ films with different thicknesses. The spectrum measured in FeTe$_{0.55}$Se$_{0.45}$ shows a fully gapped feature with a pair of coherence peaks at about $\pm1.9$ meV, similar to our previous results \cite{ChenMYCdGM}. Based on the spectra measured in Bi$_{2}$Te$_{3}$ films, one can see that the SC gapped feature is successfully induced in the Bi$_2$Te$_3$ films from the SC substrate due to the proximity effect. Shapes of spectra measured on films have traces from that measured on FeTe$_{0.55}$Se$_{0.45}$. The zero-bias differential conductance lifts with the increase of the film thickness, and the SC gap also shrinks through a simple estimation from the energy differences between two coherence peaks.

\begin{figure}
\includegraphics[width=9cm]{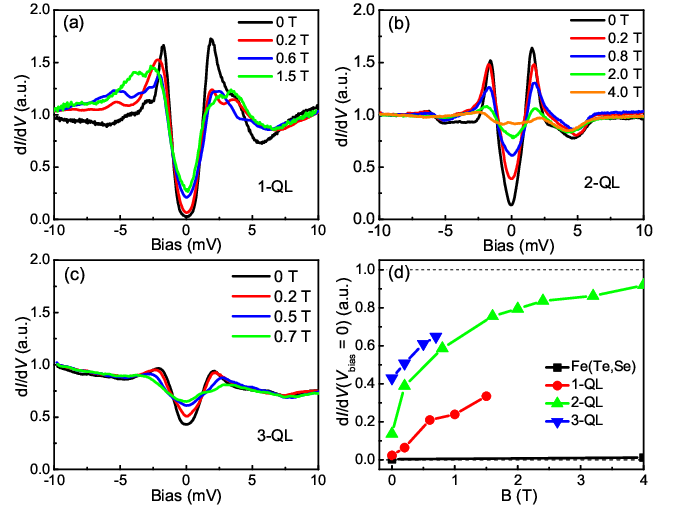}
\caption{(a)-(c) Magnetic field evolution of tunneling spectra measured on (a) 1-QL, (b) 2-QL, and (c) 3-QL films ($V_\mathrm{set} = 20$ mV, $I_\mathrm{set} = 200$ pA). These spectra are recorded at places far away from vortex cores. (d) Magnetic field dependence of the zero-bias differential conductance measured in FeTe$_{0.55}$Se$_{0.45}$ and Bi$_{2}$Te$_{3}$ films.}\label{fig2}
\end{figure}

In order to investigate the proximity-induced superconductivity in Bi$_{2}$Te$_{3}$ films, we measured the tunneling spectra under different magnetic fields. These spectra are taken at positions far away from vortex cores. The results measured on 1-QL to 3-QL films are shown in Figs.~\ref{fig2}(a) to \ref{fig2}(c), respectively. One can see that the magnetic field can lift zero-bias differential conductance easily. This effect can be seen clearly in field-dependent zero-bias differential conductance shown in Fig.~\ref{fig2}(d). However, in FeTe$_{0.55}$Se$_{0.45}$, a magnetic field of 4 T can barely affect the zero-bias differential conductance due to the extremely high upper critical field in this material. Based on the different magnetic-field-dependent behaviors of the zero-bias differential conductance, we can conclude that the field can suppress the proximity-induced superconductivity quickly in Bi$_{2}$Te$_{3}$ films.

\subsection{Vortex core and vortex bound state in 1-QL film}

\begin{figure}
\includegraphics[width=8cm]{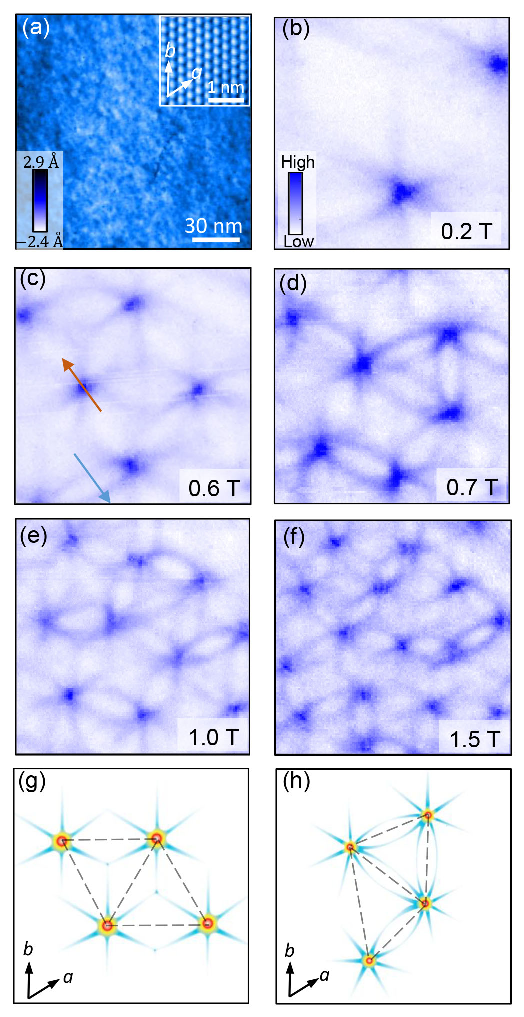}
\caption{(a) Topography measured on 1-QL Bi$_{2}$Te$_{3}$ thin film ($V_\mathrm{set} = 1$ V, $I_\mathrm{set} = 20$ pA). The inset shows the atomically resolved topography measured in this region ($V_\mathrm{set} = 10$ mV, $I_\mathrm{set} = 100$ pA), and arrows show crystalline axes of the film. (b)-(f) Zero-bias differential conductance images ($V_\mathrm{set} = 10$ mV, $I_\mathrm{set} = 200$ pA) of vortex cores measured at different magnetic fields in the same area of (a). (g),(h) Schematic images of electronic structures of two vortex lattices corresponding to local vortex lattices in (c) and (d).
} \label{fig3}
\end{figure}

Vortex cores can be imaged by the zero-bias differential conductance mapping using STM/STS measurement\cite{STMReview1,STMReview2,STMReview3}. Vortex cores on FeTe$_{0.55}$Se$_{0.45}$ are almost round-shaped \cite{ChenMYCdGM}, indicating an almost isotropic SC gap. However, in 2-QL Bi$_{2}$Te$_{3}$/FeTe$_{0.55}$Se$_{0.45}$ heterostructures, elongated vortex cores are observed, proving a twofold SC gap \cite{ChenMYSA}. This gap feature may correspond to the topological superconductivity in TI candidates \cite{ChenMYSA,NMR,Thermodynamic,LFuCuBiSe,CuBiSeSTM,NematicReview}. It should be noted that topological surface states appear only when Bi$_{2}$Te$_{3}$ films are thicker than 2 QLs \cite{BiTegrow,ChenMYSA}. Therefore, it is interesting to investigate the vortex properties in the 1-QL Bi$_{2}$Te$_{3}$/FeTe$_{0.55}$Se$_{0.45}$ heterostructure. Figures~\ref{fig3}(b)-\ref{fig3}(f) show images of vortex cores measured at different magnetic fields in the same region of Fig.~\ref{fig3}(a). At 0.2 T, one can see an isolated vortex core in the field of view of Fig.~\ref{fig3}(b). The vortex core behaves as a sixfold symmetric star shape with six rays. The rays are roughly along the crystalline axes of the Bi$_{2}$Te$_{3}$ film which are indicated in the inset of Fig.~\ref{fig3}(a). The vortex-core structure is different from that observed in FeTe$_{0.55}$Se$_{0.45}$ \cite{ChenMYCdGM} or 2-QL Bi$_{2}$Te$_{3}$/FeTe$_{0.55}$Se$_{0.45}$ heterostructures \cite{ChenMYSA}, but it is similar to the vortex-core structure measured in 2$H$-NbSe$_2$ \cite{STMReview3,Hess1,Hess2,NbS2PRL}. Theoretically, the star-shaped vortex cores in NbSe$_{2}$ can be explained by the anisotropic crystal potential \cite{vortextheory,FangDL} or the anisotropic SC gap \cite{starsplit}. The Bi$_{2}$Te$_{3}$ film has a similar sixfold symmetry of the surface lattice as 2$H$-NbSe$_{2}$, and the star-shaped core can also be explained by the in-plane sixfold SC gap or sixfold anisotropic Fermi surface. In addition, the twofold component of the vortex core is absent in the 1-QL film, in which there is also a combination of the fourfold symmetry of the substrate and the sixfold symmetry of the film. Therefore, the elongated vortex core in the 2-QL film \cite{ChenMYSA} should originate from the topological superconductivity rather than the combination of two kinds of lattice symmetries.

In 2$H$-NbSe$_{2}$, vortices arrange a perfect hexagonal lattice due to the negligible vortex pinning. The rays of the star-shaped vortex core are at $30^\circ$ to the nearest-neighbored direction of the vortex lattice, and three rays from the three nearest cores at triangle vertices intersect at the center of the unit triangle \cite{STMReview3,Hess2,NbS2PRL}. This configuration is illustrated in Fig.~\ref{fig3}(g). In the 1-QL film, we can also see such local structure, e.g., in Fig.~\ref{fig3}(c). However, when the magnetic field increases, the vortex pinning effect makes the vortex lattice disordered in the heterostructure. Some rays change their paths to the nearest neighbor direction of the vortex lattice. In this situation, a pair of rays expands from neighbor cores and forms two extra connections between these two cores. This situation is illustrated in Fig.~\ref{fig3}(h), and examples can be observed in many places in Figs.~\ref{fig3}(d) to \ref{fig3}(f). This is an interesting observation of the vortex-bound state influenced by two neighbored cores simultaneously.

\begin{figure}
\includegraphics[width=8cm]{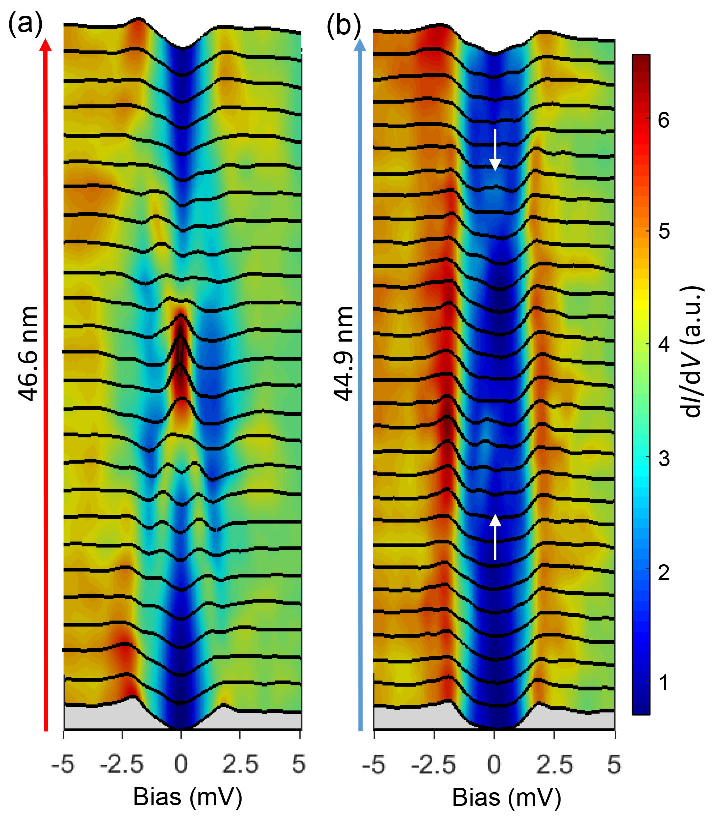}
\caption{Two sets of tunneling spectra measured across (a) a vortex core and (b) two rays connecting two neighbored cores ($V_\mathrm{set} = 20$ mV, $I_\mathrm{set} = 200$ pA). They are measured along arrows in Fig.~\ref{fig3}(c).
} \label{fig4}
\end{figure}

To study the vortex-core states and electronic properties of core rays in the 1-QL film, tunneling spectra are measured across a single vortex core and a pair of rays connecting adjacent vortices. Figure~\ref{fig4} shows these two sets of spectra measured along two arrowed lines plotted in Fig.~\ref{fig3}(c). The spatial evolution of tunneling spectra across the vortex core in Fig.~\ref{fig4}(a) is similar to the case in 2$H$-NbSe$_{2}$ \cite{STMReview3,Hess1,Hess2}. The spectrum at the center of the vortex core shows a giant peak at zero bias, and then the peak splits into two branches when the tip moves away from the center. Finally, the two peaks merge into the gap edge. This is the typical spatial evolution of vortex-bound states in a superconductor with a minimal ratio of $\Delta/E_\mathrm{F}$. Here $\Delta$ is the SC gap, and $E_\mathrm{F}$ is the Fermi energy. In this situation, the zero-bias peak or the split peaks are combined by many vortex-bound states with a negligible energy gap spacing of about $\Delta^2/E_\mathrm{F}$ \cite{vortextheory}. However, in the substrate superconductor of FeTe$_{0.55}$Se$_{0.45}$ \cite{ChenMYCdGM,DingHNP} or other iron-based superconductors \cite{ZhangTLiFe,K12442}, discrete vortex-bound states or so-called Caroli-de Gennes-Matricon (CdGM) \cite{CdGM} states are observed due to the large ratio of $\Delta/E_\mathrm{F}$. The situation of vortex-bound states in the 1-QL Bi$_{2}$Te$_{3}$ film grown on FeTe$_{0.55}$Se$_{0.45}$ is different but is like that in 2$H$-NbSe$_{2}$. In addition, the topological surface state should not exist in the 1-QL film \cite{BiTegrow,ChenMYSA}, and consistently, we do not see the noticeable feature of the zero-bias conductance peak in the vortex core center or the elongated shape of vortex cores, just like those observed in the 2-QL film \cite{ChenMYSA} or other superconductors \cite{WS2LiW,CuBiSeSTM,NematicReview}. Instead, the sixfold-symmetric vortex cores suggest a topologically trivial nature. Figure~\ref{fig4}(b) shows a set of tunneling spectra measured across a pair of rays connecting adjacent vortex cores. There is some structure near zero bias on the ray. In 2$H$-NbSe$_{2}$, the rays are caused by the lift of zero-bias conductance of the two combined vortex-bound-state peaks at positive and negative energies \cite{STMReview3}. However, the bound states inducing rays in the 1-QL Bi$_{2}$Te$_{3}$ film behave differently to those in 2$H$-NbSe$_{2}$, probably due to the curvature minima of the Fermi surface \cite{FangDL} rather than the merged vortex-bound-state peaks.

\begin{figure*}
\includegraphics[width=18cm]{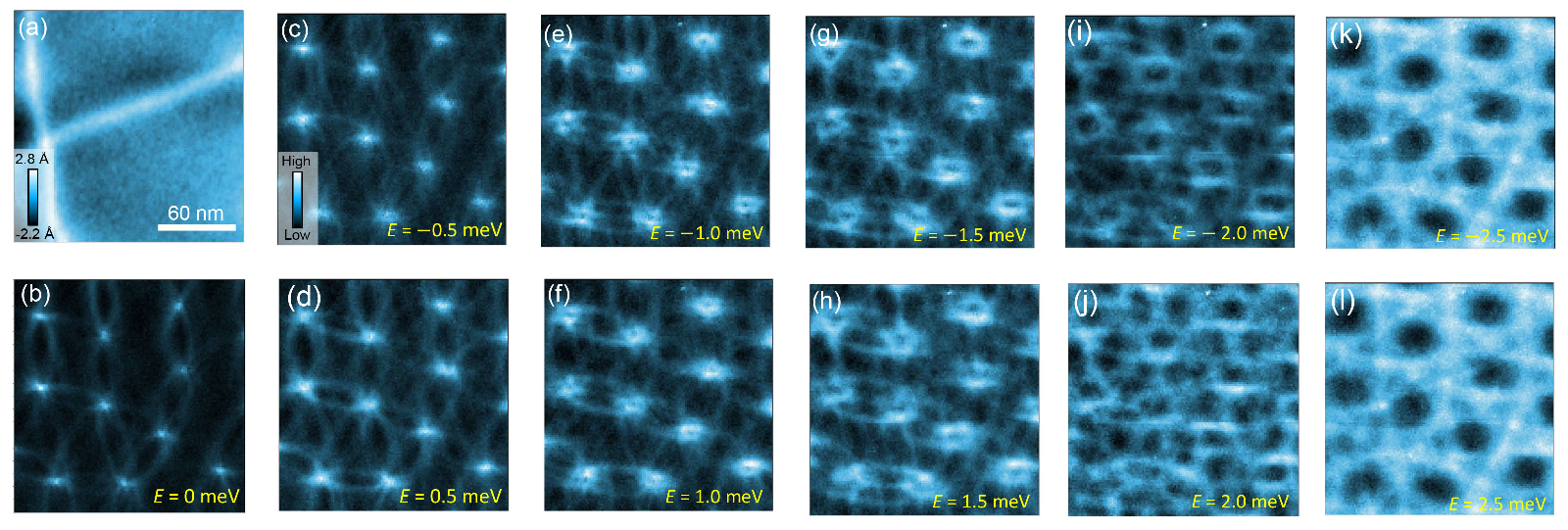}
\caption{(a) Topographic image in an area of the 1-QL film ($V_\mathrm{set} = 1$ V, $I_\mathrm{set} = 20$ pA). (b)-(l) Vortex images measured at various energies and in the same region shown in (a) ($V_\mathrm{set} = 10$ mV, $I_\mathrm{set} = 200$ pA). The applied field is 0.7 T.
} \label{fig5}
\end{figure*}

Vortex images are also recorded by doing differential-conductance mapping at different energies, and the results are shown in Fig.~\ref{fig5}. There are two ridge-like dislocations in the topography, as shown in Fig.~\ref{fig5}(a). However, they do not affect the local electronic property nor act as pinning centers to vortices, which can be derived from the differential conductance mappings in Fig.~\ref{fig5}(b). These dislocations may come from the surface strain on the substrate \cite{ChenMYSA}. In the zero-bias differential conductance mapping in Fig.~\ref{fig5}(b), images of vortex cores are similar to those shown in Fig.~\ref{fig3}(d). These vortices form the disordered lattice at 0.7 T, and pairs of rays connect adjacent vortex cores. With the increase of energy (Figs.~\ref{fig5}(c)-\ref{fig5}(h)), the ray seems to split into a double branch. This is similar to the situation in the 2-QL film and can be explained theoretically from the spatial and energy evolution of vortex-bound states \cite{starsplit,gapanisotropy}. With further energy increase (Figs.~\ref{fig5}(i)-\ref{fig5}(l)), the dark round discs appear in the center of vortex cores, indicating a sightly suppression of the density of states within the discs. This is similar to the vortex cores measured at high energies in the 2-QL film \cite{ChenMYSA}.

\subsection{Vortex images and vortex bound states on 3-QL heterostructures}

\begin{figure}[b]
\includegraphics[width=8cm]{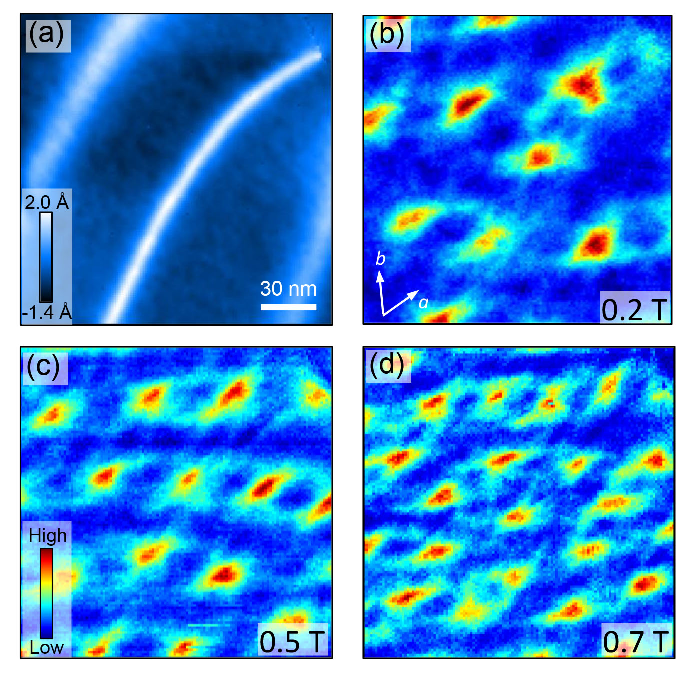}
\caption{(a) Topographic image in an area on the 3-QL film ($V_\mathrm{set} = 1$ V, $I_\mathrm{set} = 10$ pA). (b)-(d) Vortex images recorded at zero bias under different magnetic fields ($V_\mathrm{set} = 10$ mV, $I_\mathrm{set} = 200$ pA). Vortex cores are elongated along one pair of the crystalline axes.
} \label{fig6}
\end{figure}

In the 1-QL Bi$_{2}$Te$_{3}$ film, the vortex cores manifest as a sixfold shape, which suggests a topologically trivial superconductivity. In contrast, we observed elongated vortices in the 2-QL film. As mentioned above, the sixfold shape of the vortex core in the 1-QL film rules out the possibility that the elongated vortex core in 2-QL film \cite{ChenMYSA} originates from the combination of the substrate's fourfold symmetry and the film's sixfold symmetry. Therefore, the elongated vortex core is evidence of the twofold symmetry of the SC gap and the existence of the topological nontrivial superconductivity \cite{ChenMYSA,NMR,Thermodynamic,LFuCuBiSe,CuBiSeSTM,NematicReview}. Since topological surface states appear when Bi$_{2}$Te$_{3}$ film is thicker than 2 QLs \cite{BiTegrow,ChenMYSA}, it is worth imaging the vortex cores on 3-QL film to see whether the twofold symmetric feature still exists. Figure~\ref{fig6}(a) shows the topography measured in a 3-QL film, and one can see some ridge-like dislocations. Vortex cores are imaged by the zero-bias differential conductance mapping, and the results under different fields are presented in Figs.~\ref{fig6}(b)-\ref{fig6}(d). Vortices are not pinned by the ridge-like dislocations on the topography. Moreover, vortex cores are all elongated along one of the crystalline axes (defined as the $a$ axis) of the film, similar to the result on the 2-QL film \cite{ChenMYSA}. Thus, we can infer that the SC gap is also twofold symmetric in 3-QL heterostructures, and the topological superconductivity is likely to exist in the TI film of this thickness.

\begin{figure}
\includegraphics[width=8.5cm]{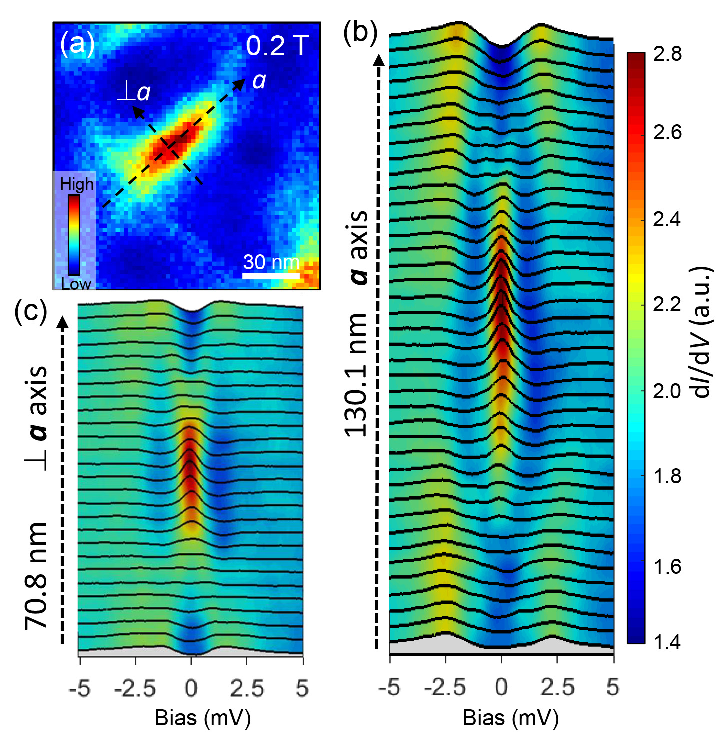}
\caption{(a) Image of an isolated vortex core measured at zero bias, and it is elongated along one pair of the crystalline axes which we can define as the $a$ axis. Two sets of tunneling spectra measured (b) along and (c) perpendicular to the $a$-axis direction, respectively ($V_\mathrm{set} = 10$ mV, $I_\mathrm{set} = 100$ pA).
} \label{fig7}
\end{figure}

In order to study the vortex-bound states in the vortex cores on the 3-QL Bi$_{2}$Te$_{3}$ film, we also measure tunneling spectra along and perpendicular to the vortex elongating directions ($a$ axis), and results are shown in Fig.~\ref{fig7}. Along both directions, spatial evolution of vortex-bound states behaves like that in a typical superconductor with large $E_\mathrm{F}$ \cite{STMReview3,Hess1,Hess2,vortextheory}. However, the distance with a zero-bias conductance peak seems to be longer than normal vortex core, which is similar to the situation in other topological-superconductor candidates \cite{Jiaprl,ChenMYSA,WS2LiW,CVSWangZY}. Therefore, there may be Majorana zero mode coexisting with conventional vortex bound states in the vortex core in the 3-QL film.

\section{Discussion}

\begin{figure}
\centering
\includegraphics[width=9cm]{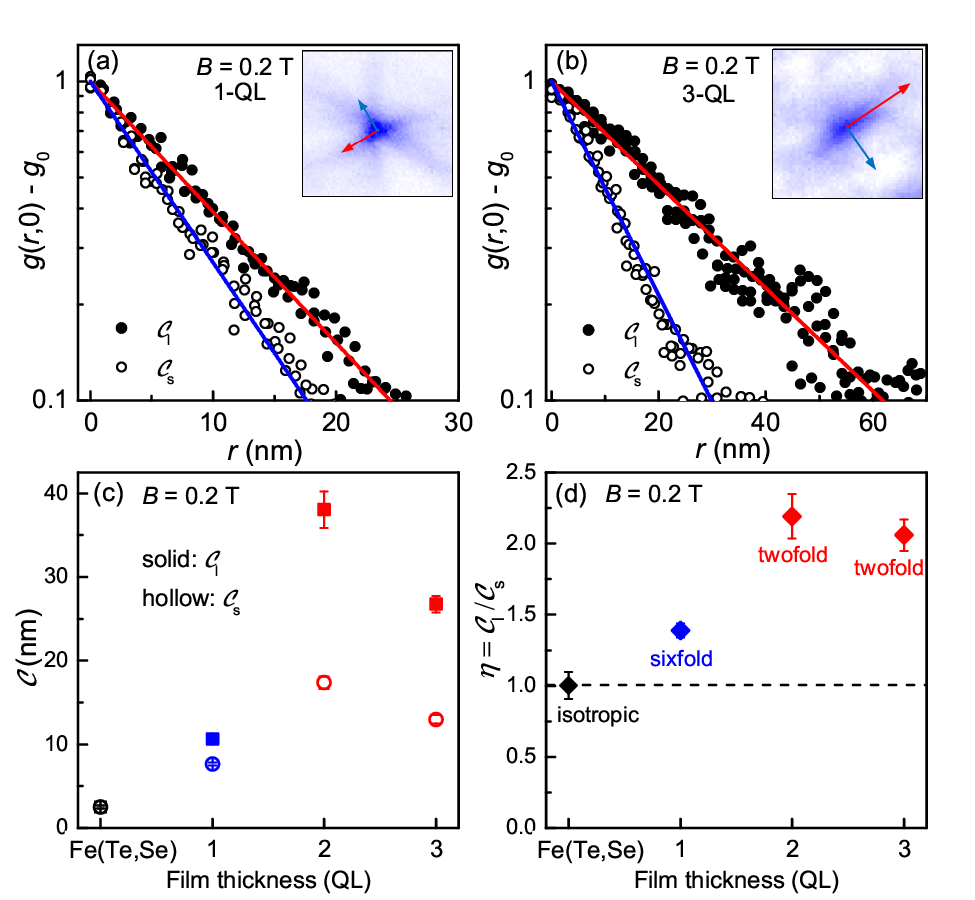}
\caption{Spatial dependence of the normalized differential conductance at zero bias $g(r,0)$ in semi-log plots across several different vortices along the longer and the shorter axes of vortex cores in (a) the 1-QL and (b) the 3-QL film. The differential conductance $g_0$ measured extremely far away from the core center has been subtracted as a background. The direction with the larger core size is along the ray in the 1-QL film (inset of (a)) or along the elongating direction in the 3-QL film (inset of (b)), and both of these directions are supposed to be roughly along the $a$ axis of the film. The longer ($\mathcal{C}_\mathrm{l}$) and shorter ($\mathcal{C}_\mathrm{s}$) vortex sizes with error bars are obtained by fitting to the experimental data by the exponential decay formula. (c) Thickness dependent $\mathcal{C}_\mathrm{l}$ and $\mathcal{C}_\mathrm{s}$ derived from the fitting results. (d) Thickness dependent anisotropy of the core size ($\eta=\mathcal{C}_\mathrm{l}/\mathcal{C}_\mathrm{s}$).
} \label{fig8}
\end{figure}

Vortex-core images and vortex-bound states have been investigated in Bi$_{2}$Te$_{3}$/FeTe$_{0.55}$Se$_{0.45}$ heterostructures with different thicknesses of Bi$_{2}$Te$_{3}$ layers. The sixfold symmetric star-shaped vortices are observed in the 1-QL film, which can be explained by the topologically trivial nature when considering a sixfold symmetry of the Fermi surface \cite{vortextheory,FangDL} or an SC gap with the sixfold symmetry \cite{starsplit}. There are six rays along crystalline axes around an isolated vortex core. While in the heterostructures with Bi$_{2}$Te$_{3}$ film thicker than 2 QLs, elongated vortices have been observed, and the elongated direction is along one pair of crystalline axes. This kind of vortex core can be explained by the twofold symmetric $\Delta_{4y}$ SC gap function proposed for topological superconductors \cite{LFuCuBiSe}. Therefore, images of vortex cores are anisotropic in heterostructures with Bi$_{2}$Te$_{3}$-film thicknesses of more than 2 QLs. To get the anisotropy of the core size, we fit the spatial ($r$) dependence of the differential conductance at zero bias ($g(r,0)$) by using the exponential decay formula $g(r,0) = g_{0} + Ae^{-r/\mathcal{C}}$ \cite{HoffmanPRL,TaPdTe} with $g_{0}$ the background signal obtained far away from the vortex core and $\mathcal{C}$ the core size. We focus on vortex cores imaged at 0.2 T to avoid the influence of neighboring vortex cores. The fitting results to the 1-QL and 3-QL films are shown in Figs.~\ref{fig8}(a) and \ref{fig8}(b) as examples, and then the longer ($\mathcal{C}_\mathrm{l}$) and the shorter ($\mathcal{C}_\mathrm{s}$) core size can be obtained and they are shown in Fig.~\ref{fig8}(c).

The core size $\mathcal{C}$ may roughly equal to the coherence length $\xi$. For example, the core size is about 2.58 nm for the substrate FeTe$_{0.55}$Se$_{0.45}$ from fitting the experimental data in Ref.~\cite{ChenMYCdGM}, consistent with the coherence length of 2.64 nm derived from the upper critical field of about 47 T \cite{coherencelength}. Then the anisotropy of $\mathcal{C}$ or $\xi$ may be due to the anisotropy of the SC gap based on the formula $\xi={\hbar v_{F}}/{\pi \Delta}$ with $v_{F}$ the Fermi velocity \cite{gapanisotropy}. The core size anisotropy $\eta$ is calculated from the definition of $\eta=\mathcal{C}_\mathrm{l}/\mathcal{C}_\mathrm{s}$, and values are shown in Fig.~\ref{fig8}(d). $\eta$ is about 1.0 for vortex cores in FeTe$_{0.55}$Se$_{0.45}$, suggesting an isotropic feature of the vortex core in the substrate superconductor. In the 1-QL film, although the rays seem to be very long, its zero-bias density of states is barely enhanced. Therefore, the anisotropy of the sixfold vortex core is only about 1.4 in the 1-QL film. In 2-QL and 3-QL films, the anisotropy of elongated vortices is about 2.2, which is the anisotropy of the SC gap in 2-QL and 3-QL films if the Fermi surface is almost isotropic.

The vortex-bound states on FeTe$_{0.55}$Se$_{0.45}$ or Bi$_{2}$Te$_{3}$/FeTe$_{0.55}$Se$_{0.45}$ heterostructures are also different. Discrete CdGM bound-states are observed in some vortex cores in FeTe$_{0.55}$Se$_{0.45}$ because of the minimal Fermi energy \cite{ChenMYCdGM}. However, the vortex-bound states on 1-QL Bi$_{2}$Te$_{3}$ films coincide with the merged CdGM states due to its large $E_F$, i.e., a giant zero-bias peak appears at the center of the vortex core and then it splits into two branches when the tip moves away from the vortex center. The change of the vortex-bound state may originate from a dramatic increase of Fermi energy on the Bi$_{2}$Te$_{3}$ surface, no matter whether the proximity-induced superconductivity is topologically nontrivial or trivial in the Bi$_{2}$Te$_{3}$ film. The considerable change of the vortex core shape and bound states with different film thicknesses of 1 and 2 QLs may originate from topological surface states on 2-QL or thicker Bi$_{2}$Te$_{3}$ films. In 2- and 3-QL Bi$_{2}$Te$_{3}$ films, the vortex-bound states along the elongated direction show a zero-bias peak which persists a relatively long distance, and this feature may be induced by the mixture of the Majorana zero mode and conventional CdGM states.

\section{Conclusion}

To conclude, we investigate electronic and vortex properties in Bi$_{2}$Te$_{3}$/FeTe$_{0.55}$Se$_{0.45}$ heterostructures with different thicknesses of the Bi$_{2}$Te$_{3}$ film. The gapped feature and coherence peaks on the tunneling spectra reveal that superconductivity is successfully induced to the Bi$_{2}$Te$_{3}$ thin films from the iron-based superconductor FeTe$_{0.55}$Se$_{0.45}$. The sixfold-symmetric star-shaped vortices with extended rays indicate the sixfold symmetry of the Fermi surface or the SC gap on 1-QL heterostructures. An interesting connection of neighbored vortex cores by a pair of rays is observed when the arrangement of vortices deviates from the standard triangular vortex lattice. Being similar to those in the 2-QL heterostructure, elongated vortex cores are observed on the 3-QL heterostructure, indicating a twofold symmetric and possibly topologically nontrivial superconductivity in these heterostructures. Our observations will shed some valuable insights into understanding the proximity-induced topological superconductivity.

\begin{acknowledgments}
The work was supported by the National Natural Science Foundation of China (Grants No. 11974171, No. 12061131001, and No. 11927809) and the National Key R\&D Program of China (Grant No. 2022YFA1403201).
\end{acknowledgments}

$^*$ huanyang@nju.edu.cn

$^\dag$ hhwen@nju.edu.cn


\begin{thebibliography}{40}

\bibitem{SCZhangReview} X. L. Qi and S. C. Zhang, Rev. Mod. Phys. \textbf{83}, 1057 (2011).

\bibitem{AndoReview} M. Sato and Y. Ando, Rep. Prog. Phys. \textbf{80}, 076501 (2017).

\bibitem{CuBiSe1} Y. S. Hor, A. J. Williams, J. G. Checkelsky, P. Roushan, J. Seo, Q. Xu, H. W. Zandbergen, A. Yazdani, N. P. Ong, and R. J. Cava, Phys. Rev. Lett. \textbf{104}, 057001 (2010).

\bibitem{CuBiSe2} L. A. Wray, S.-Y. Xu, Y. Xia, Y.S. Hor, D. Qian, A. V. Fedorov, H. Lin, A. Bansil, R. J. Cava, and M. Z. Hasan, Nat. Phys. \textbf{6}, 855 (2010).

\bibitem{CuBiSe3} N. Levy, T. Zhang, J. Ha, F. Sharifi, A. A. Talin, Y. Kuk, and J. A. Stroscio, Phys. Rev. Lett. \textbf{110}, 117001 (2013).

\bibitem{SrBiSe} Z. Liu, X. Yao, J. Shao, M. Zuo, L. Pi, S. Tan, C. Zhang, and Y. Zhang, J. Am. Chem. Soc. \textbf{137}, 10512 (2015).

\bibitem{GDu} G. Du, J. Shao, X. Yang, Z. Du, D. Fang, J. Wang, K. Ran, J. Wen, C. Zhang, H. Yang, Y. Zhang, and H.-H. Wen, Nat. Commun. \textbf{8}, 14466 (2017).

\bibitem{HuReview} N. Hao and J. Hu, Natl. Sci. Rev. \textbf{6}, 213 (2018).

\bibitem{FeTeSeMajorana} D. F. Wang, L. Y. Kong, P. Fan, H. Chen, S. Y .Zhu, W. Y. Liu, L. Cao, Y. J. Sun, S. X. Du, J. Schneeloch, R. D. Zhong, G. D. Gu, L. Fu, H. Ding, and H.-J. Gao, Science \textbf{362}, 6412 (2018).

\bibitem{FeTeSearpes} P. Zhang, K. Yaji, T. Hashimoto, Y. Ota, T. Kondo, K. Okazaki, Z. Wang, J. Wen, G. D. Gu, H. Ding, and S. Shin, Science \textbf{360}, 182 (2018).

\bibitem{LiFeOHFeSe} Q. Liu, C. Chen, T. Zhang, R. Peng, Y. J. Yan, C. H. P. Wen, X. Lou, Y. L. Huang, J. P. Tian, X. L. Dong, G. W. Wang, W. C. Bao, Q. H. Wang, Z. P. Y, Z. X. Zhao, and D. L. Feng, Phys. Rev. X \textbf{8}, 041056 (2018).

\bibitem{FeTeSeHanaguri} T. Machida, Y. Sun, S. Pyon, S. Takeda, Y. Kohsaka, T. Hanaguri, T. Sasagawa, and T. Tamegai, Nat. Mater. \textbf{18}, 811 (2019).

\bibitem{CaKFe4As4} W. Liu, L. Cao, S. Zhu, L. Kong, G. Wang, M. Papaj, P. Zhang, Y.-B. Liu, H. Chen, G. Li, F. Yang, T. Kondo, S. Du, G.-H. Cao, S. Shin, L. Fu, Z. Yin, H.-J. Gao, and H. Ding, Nat. Commun. \textbf{11}, 5688 (2020).

\bibitem{WangZY} Z. Wang, J. O. Rodriguez, L. Jiao, S. Howard, M. Graham, G. D. Gu, T. L. Hughes, D. K. Morr, and V. Madhavan, Science \textbf{367}, 104 (2020)

\bibitem{LiFeAs} M. Li, G. Li, L. Cao, X. Zhou, X. Wang, C. Jin, C.-K. Chiu, S. J. Pennycook, Z. Wang, and H.-J. Gao, Nature (London) \textbf{606}, 890 (2022).

\bibitem{WS2LiW} Y. Yuan, J. Pan, X. Wang, Y. Fang, C. Song, L. Wang, K. He, X. Ma, H. Zhang, F. Huang, W. Li, and Q.-K. Xue, Nat. Phys. \textbf{15}, 1046 (2019).

\bibitem{CVSWangZY} Z. Liang, X. Hou, F. Zhang, W. Ma, P. Wu, Z. Zhang, F. Yu, J.-J. Ying, K. Jiang, L. Shan, Z. Wang, and X.-H. Chen, Phys. Rev. X \textbf{11}, 031026 (2021).

\bibitem{Jiaolin} L. Jiao, S. Howard, S. Ran, Z. Wang, J. O. Rodriguez, M. Sigrist, Z. Wang, N. P. Butch, and V. Madhavan, Nature (London) \textbf{579}, 523 (2020).

\bibitem{LFuProximity} L. Fu and C. L. Kane, Phys. Rev. Lett. \textbf{100}, 096407 (2008).

\bibitem{JiaReview} H.-H. Sun and J.-F. Jia, npj Quantum Mater. \textbf{2}, 34 (2017).

\bibitem{Jiaprl} J.-P. Xu, M.-X. Wang, Z. L. Liu, J.-F. Ge, X. Yang, C. Liu, Z. A. Xu, D. Guan, C. L. Gao, D. Qian, Y. Liu, Q.-H. Wang, F.-C. Zhang, Q.-K. Xue, and J.-F. Jia, Phys. Rev. Lett. \textbf{114}, 017001 (2015).

\bibitem{FeChain} S. Nadj-Perge, I. K. Drozdov, J. Li, H. Chen, S. Jeon, J. Seo, A. H. MacDonald, B. A. Bernevig, and A. Yazdani, Science \textbf{346}, 602 (2014).

\bibitem{Feisland} A. Palacio-Morales, E. Mascot, S. Cocklin, H Kim, S. Rache, D. K. Morr, and R. Wiesendanger, Sci. Adv. \textbf{5}, eaav6600 (2019).

\bibitem{CrBr3} S. Kezilebieke, M. N. Huda, V. Va\u{n}o, M. Aapro, S. C. Ganguli, O. J. Silveira, S. G{\l}odzik, A. S. Foster, T. Ojanen, and P. Liljeroth, Nature (London) \textbf{588}, 424 (2020).

\bibitem{TIonBi2212} P. Zareapour, A. Hayat, S. Y. F. Zhao, M. Kreshchuk, A. Jain, D. C. Kwok, N. Lee, S. W. Cheong, Z. Xu, A. Yang, G. D. Gu, S. Jia, R. J. Cava, and K. S. Burch, Nat. Commun. \textbf{3}, 1056 (2012).

\bibitem{TIonBi2212ARPES} E. Wang, H. Ding, A. V. Fedorov, W. Yao, Z. Li, Y.-F. Lv, k. Zhao, L.-G. Zhang, Z. Xu, J. Schneeloch, R. Zhong, S.-H. Ji, L. Wang, K. He, X. Ma, G. Gu, H. Yao, Q. K. Xue, X. Chen, and S. Zhou, Nat. Phys. \textbf{9}, 621 (2013).

\bibitem{TIonBi2212Wan} S. Wan, Q. Gu, H. Li, H. Yang, J. Schneeloch, R. D. Zhong, G. D. Gu, and H.-H. Wen, Phys. Rev. B \textbf{101}, 220503(R) (2020).

\bibitem{LFuCuBiSe} L. Fu, Phys. Rev. B \textbf{90}, 100509(R) (2014).


\bibitem{ChenMYSA} M. Chen, X. Chen, H. Yang, Z. Du, and H.-H. Wen, Sci. Adv. \textbf{4}, eaat1084 (2018).

\bibitem{BiTeonFeTeSe} H. Zhao, B. Rachmilowitz, Z. Ren, R. Han, J. Schneeloch, R. D. Zhong, G. D. Gu, Z. Q. Wang, and I. Zeljkovic, Phys. Rev. B \textbf{97}, 224504 (2018).

\bibitem{Biisland} X. Chen, M. Chen, W. Duan, H. Yang, and H.-H. Wen, Nano Lett. \textbf{20}, 2965 (2020).

\bibitem{MnTe} S. Ding, C. Chen, Z. Cao, D. Wang, Y. Pan, R. Tao, D. Zhao, Y. Hu, T. Jiang, Y. Yan, Z. Shi, X. Wan, D. Feng, and T. Zhang, Sci. Adv. \textbf{8}, eabq4578 (2022).


\bibitem{samplegrowth} Y. Liu and C. T. Lin, J. Supercond. Nov. Magn. \textbf{24}, 183 (2011).

\bibitem{ChenMYCdGM} M. Chen, X. Chen, H. Yang, Z. Du, X. Zhu, E. Wang, and H.-H. Wen, Nat. Commun. \textbf{9}, 970 (2018).

\bibitem{STMReview1} {\O}. Fischer, M. Kugler, I. Maggio-Aprile, C. Berthod, and C. Renner, Rev. Mod. Phys. \textbf{79}, 353 (2007).

\bibitem{STMReview2} J. E. Hoffman, Rep. Prog. Phys. \textbf{74}, 124513 (2011).

\bibitem{STMReview3} H. Suderow, I. Guillam\'{o}n, J. G. Rodrigo, and S. Vieira, Supercond. Sci. Technol. \textbf{27}, 063001 (2014).

\bibitem{NMR} K. Matano, M. Kriener, K. Segawa, Y. Ando, and G.-Q. Zheng, Nat. Phys. \textbf{12}, 852 (2016).

\bibitem{Thermodynamic} S. Yonezawa, K. Tajiri, S. Nakata, Y. Nagai, Z. Wang, K. Segawa, Y. Ando, and Y. Maeno, Nat. Phys. \textbf{13}, 123 (2017).

\bibitem{CuBiSeSTM} R. Tao, Y.-J. Yan, X. Liu, Z.-W. Wang, Y. Ando, Q.-H. Wang, T. Zhang, and D.-L. Feng, Phys. Rev. X \textbf{8}, 041024 (2018).

\bibitem{NematicReview} S. Yonezawa, Condens. Matter \textbf{4}, 2 (2019).

\bibitem{BiTegrow} Y. Y. Li, G. Wang, X. G. Zhu, M. H. Liu, C. Ye, X. Chen, Y. Y. Wang, K. He, L. L. Wang, X. C. Ma, H. J. Zhang, X. Dai, Z. Fang, X. C. Xie, Y. Liu, X. L. Qi, J. F. Jia, S. C. Zhang, and Q. K. Xue, Adv. Mater. \textbf{22}, 4002 (2010).

\bibitem{Hess1} H. F. Hess, R. B. Robinson, and J. V. Waszczak, Phys. Rev. Lett. \textbf{64}, 2711 (1990).

\bibitem{Hess2} H. F. Hess, Physica (Amsterdam) \textbf{185-189C}, 259 (1991).

\bibitem{NbS2PRL} I. Guillam\'{o}n, H. Suderow, S. Vieira, L. Cario, P. Diener, and P. Rodi\`{e}re, Phys. Rev. Lett. \textbf{101}, 166407 (2008).

\bibitem{vortextheory} F. Gygi and M. Schl\"{u}ter, Phys. Rev. B \textbf{43}, 7609 (1991).

\bibitem{FangDL} D. Fang, Chin. Phys. B \textbf{32}, 037403 (2023).

\bibitem{starsplit} N. Hayashi, M. Ichioka, and K. Machida, Phys. Rev. Lett. \textbf{77}, 4074 (1996).

\bibitem{DingHNP} L. Kong, S. Zhu, M. Papaj, L. Cao, H. Isobe, W. Liu, D. Wang, P. Fan, H. Chen, Y. Sun, S. Du, J. Schneeloch, R. Zhong, G. Gu, L. Fu, H.-J. Gao, and H. Ding, Nat. Phys. \textbf{15}, 1181 (2019).

\bibitem{ZhangTLiFe} T. Zhang, W. Bao, C. Chen, D. Li, Z. Lu, Y. Hu, W. Yang, D. Zhao, Y. Yan, X. Dong, Q.-H. Wang, T. Zhang, and D. Feng, Phys. Rev. Lett. \textbf{126}, 127001 (2021).

\bibitem{K12442} X. Chen, W. Duan, X. Fan, W. Hong, K. Chen, H. Yang, S. Li, H. Luo, and H.-H. Wen, Phys. Rev. Lett. \textbf{126}, 257002 (2021).

\bibitem{CdGM} C. Caroli, P. G. De Gennes, and J. Matricon, Phys. Lett. \textbf{9}, 307 (1964).

\bibitem{gapanisotropy} N. Hayashi, M. Ichioka, and K. Machida, Phys. Rev. B \textbf{56}, 9052 (1997).

\bibitem{HoffmanPRL} Y. Yin, M. Zech, T. L. Williams, X. F. Wang, G. Wu, X. H. Chen, and J. E. Hoffman, Phys. Rev. Lett. \textbf{102}, 097002 (2009).

\bibitem{TaPdTe} Z. Du, D. Fang, Z. Wang, Y. Li, G. Du, H. Yang, X. Zhu, and H.-H. Wen, Sci. Rep. \textbf{5}, 9408 (2015).

\bibitem{coherencelength} D. Braithwaite, G. Lapertot, W. Knafo, and I. Sheikin, J. Phys. Soc. Jpn. \textbf{79}, 053703 (2010).

\end{thebibliography}
\end{document}